# Raman fingerprints of atomically precise graphene nanoribbons

Ivan A. Verzhbitskiy,[1,†] Marzio De Corato,[2,3] Alice Ruini,[2,3] Elisa Molinari,[2,3] Akimitsu Narita,[4] Yunbin Hu,[4] Matthias Georg Schwab,[4,‡] M. Bruna,[5] D. Yoon,[5] S. Milana,[5] Xinliang Feng,[6] Klaus Müllen,[4] Andrea C. Ferrari,[5] Cinzia Casiraghi,[1,7,*] and Deborah Prezzi[3,*]

[1] *Physics Department, Free University Berlin, Germany*

[2] *Dept. of Physics, Mathematics, and Informatics, University of Modena and Reggio Emilia, Modena, Italy*

[3] *Nanoscience Institute of CNR, S3 Center, Modena, Italy*

[4] *Max Planck Institute for Polymer Research, Mainz, Germany*

[5] *Cambridge Graphene Centre, University of Cambridge, Cambridge, CB3 0FA, UK*

[6] *Center for Advancing Electronics Dresden (cfaed) & Department of Chemistry and Food Chemistry, Technische Universitaet Dresden, Germany*

[7] *School of Chemistry, University of Manchester, UK*

[†] Present address: Physics Department, National University of Singapore, 117542, Singapore

[‡] Present address: BASF SE, Carl-Bosch-Straße 38, 67056 Ludwigshafen, Germany

* Corresponding authors: (DP) deborah.prezzi@nano.cnr.it; (CC) cinzia.casiraghi@manchester.ac.uk

**Bottom-up approaches allow the production of ultra-narrow and atomically precise graphene nanoribbons (GNRs), with electronic and optical properties controlled by the specific atomic structure. Combining Raman spectroscopy and ab-initio simulations, we show that GNR width, edge geometry and functional groups all influence their Raman spectra. The low-energy spectral region below 1000 cm$^{-1}$ is particularly sensitive to edge morphology and functionalization, while the D peak dispersion can be used to uniquely fingerprint the presence of GNRs, and differentiates them from other sp$^2$ carbon nanostructures.**

Raman spectroscopy is one of the most used characterization techniques in carbon science and technology.[1–7] In the last decade, the rise of graphene have triggered extensive Raman studies to understand not only phonons, but also electron-phonon, magneto-phonon and electron-electron interactions, as well as the influence of the number and orientation of the layers in few-layer graphene (FLG), electric or magnetic fields, strain, doping, disorder, quality and type of edges, and functional groups.[1] The Raman spectrum of single-layer graphene (SLG) and FLG consists of two sets of peaks: the so-called G and D peaks, which originate from in-plane vibrations and dominate the optical region of graphene and other $sp^2$ bonded materials;[8,9] the low-energy peaks resulting from the relative motion of the planes, such as the shear[10] (C) and the layer-breathing modes[11–13] (LBMs), which can be used as a direct probe of the number of layers.

The low-energy region, where the C and LBMs are located, is particularly interesting because specific fingerprints have been observed for a variety of carbon allotropes. Most notably, in the case of carbon nanotubes (CNTs), a characteristic peak at low energy, associated to the radial breathing mode (RBM) of all the atoms of the structure, has been widely exploited to determine the tube diameter.[14] Polycyclic aromatic hydrocarbons (PAHs) are also characterized by breathing-like low-energy modes that can be related to their lateral size.[15,16] GNRs –of interest for emerging digital nanoelectronics, optoelectronics and spintronics[17–19]— are also expected to show characteristic Raman features in this spectral region due to their finite width and low symmetry.[20–25]

Recently, techniques based on surface-assisted[26] or solution-based[27–29] cyclo-dehydrogenation of tailor-made aromatic polymer precursors have enabled the production of well-defined GNRs, with lateral width well below 10 nm and defined edge patterns.[26–31] In the case of GNRs prepared by solution-based processing, the edges are typically functionalized by alkyl chains to improve solubility.[29] Thus, these GNRs provide a unique opportunity to study the combined effect of finite width and edge patterns, and gain insights into the evolution of the vibrational properties with lateral size. At the same time, such GNRs call for the identification of specific Raman fingerprints, which could be extremely relevant for their characterization and design, similar to what happened in the case of CNTs.

In this respect, experiments on GNRs mostly targeted peaks in the D/G energy region.[27,31–33] However, these features are usually just discussed as a hallmark of $sp^2$ hybridization, and no further analysis exists so far. A few groups studied the low-energy GNR Raman bands. Cai et al.[26] reported a sharp (20-30 cm$^{-1}$) low-energy peak at ~400 cm$^{-1}$, and assigned it to a Radial-Like Breathing Mode[20] (RLBM), where all the atoms of the ribbon move in-plane along to the ribbon width direction, in analogy to the RBM in CNTs. Because of that, the position of this peak is determined by the GNR width only.[20–25] A similar feature was reported in Refs. [29,30] for other GNRs produced by solution-based processing, but much broader (~100 cm$^{-1}$ in width) when compared to Ref. [26], with several sub-components, and downshifted of ~50 cm$^{-1}$ compared to what expected from the inverse dependence of the RLBM frequency with GNR width.[20–25] This may indicate a more complex dependence of the RLBM on GNR width and edge type, and call for further investigations.

Here, we report a combined experimental and theoretical study of the Raman spectrum of ultra-narrow (< 2nm) GNRs with well-defined edges, produced by solution-based methods. By investigating four GNRs with different width, edge-pattern and location of the alkyl groups, we demonstrate that the RLBM can be influenced by the edge termination, and not just by the width. We also show that the D peak dispersion with excitation energy is a further fingerprint of these GNRs, and uniquely differentiates them from other sp$^2$ carbon systems, such as graphene or CNTs.

We investigate four GNRs, which are functionalized with long alkyl chains (–$C_{12}H_{25}$) to favor their solubility.[30] The structures in Figure 1a-c represent a series of "cove-shaped" GNRs with the same edge pattern and increasing width, where a benzo ring periodically decorates the zigzag (ZZ) edge. Following Ref. [34], these structures are labeled $n$CNRs, where $n$ indicates the width of the ZZ GNR core (here $n$ = 4, 6, 8). In addition to those samples, we also investigate a GNR based on the chiral-edged (4,1)-GNR with periodically fused benzo rings. This is called $m$-ANR in Ref. [34] and here shown in Figure 1d. The $m$-ANR has the same width as the 8CNR, but its "chirality" leads to slightly different edge pattern. The 6CNR has intermediate width between 4CNR and 8CNR and the same edge pattern, but the alkyl groups are placed in a different location at the edge: 4- and 8CNR are substituted at the outer positions on the benzo rings, while the 6CNR is substituted at the inner position inside the cove-type edge.

**Low-energy spectral region and radial-like breathing mode.** Figure 2a shows the low-energy Raman spectrum of the GNRs in Figure 1. In the case of 4CNR, one broad peak appears at ~230 cm$^{-1}$ (Full Width at Half Maximum, FWHM ~100 cm$^{-1}$), which comprises at least three contributions. For 6CNR, 8CNR and $m$-ANR, the most intense peak is located at ~130-150 cm$^{-1}$, and its width is sharp for $m$-ANR (~30 cm$^{-1}$), while it is larger for 8CNR and 6CNR (~80-100 cm$^{-1}$).

Furthermore, depending on the exact GNR structure, the peak disappears at certain excitation energies. For instance, the peak is not seen at energies below 1.96 eV for 4CNR, while for 8CNR it is detected at 1.96 eV. Other peaks with small intensity are also seen in the spectral range 400-700 cm$^{-1}$, as shown in Figure 2a. This is different from the case of CNTs and armchair GNRs grown on substrate,[26] where the low energy region is dominated by a sharp breathing mode, with frequency uniquely determined by the CNT/GNR lateral size and described by the zone-folding (ZF) approximation.[24,25]

In order to gain further insights into the origin of the GNRs Raman peaks at low energy, we have computed from first principles the Raman spectra of several GNRs (see Methods), by increasing the structure complexity in a stepwise manner. We first consider perfect ZZ-GNRs, corresponding to the cores of the cove-shaped ribbons studied here (indicated by the shaded areas in Figure 1a-c). We then study cove-shaped GNRs with hydrogen-terminated edges. Finally, we compare the above systems with functionalized ones, as in the real samples, though simulated by using shorter alkyl chains (–$C_4H_9$ instead of –$C_{12}H_{25}$) to make the calculations more affordable. Figure 3a shows the simulated Raman spectrum of the 8CNR, while the spectra of the other GNRs are reported in Supplementary Information. Let us first focus on the H-passivated 8CNR, hereafter labeled 8CNR+H. The spectrum is characterized by a dominant peak at 183 cm$^{-1}$, which falls in-between the RLBMs of the ZZ-GNRs corresponding to the minimum and maximum width of this cove-shaped GNR, i.e. 209 and 164 cm$^{-1}$ for the 8- and 10-ZZ GNRs, respectively. This behavior, common to all the GNRs studied here, can be understood in terms of an effective width model. Indeed, these cove-shaped GNRs present a modulated structure, with variable width (e.g. 4CNR core width ranges from 0.69nm to 1.13nm). However, we can define an effective width as the weighted average of the different GNR widths (see Supplementary Information), and compare our first principles simulations with the results obtained by ZF. Figure 3b shows that the agreement is within less than 10 cm$^{-1}$ for GNRs wider than 15 Å, as expected for this approximation.[24,25] Besides the RLBM, there are higher frequency modes (470-680 cm$^{-1}$), combining longitudinal and transverse components: these, not present in ZZ-GNRs, appear in cove-shaped GNRs in view of the different periodicity along the ribbon axis introduced by the additional benzo rings at the edge.

We next functionalize the edges with short alkyl chains (8CNR + $C_4H_9$). We observe a further red-shift of the RLBM peak, which moves from 183 to 156 cm$^{-1}$, close to that of the larger ZZ component of the 8CNR, i.e. 10-ZZ GNR, even though the origin of the shift is rather different, as clarified by our analysis. The breathing does not involve the ribbon only, but also part of the chain that moves in phase with the ribbon atoms. This causes a redshift as a result of an increase in the effective width. We further check the effect of the chains by varying the chain length, as shown in Figure 3d. In addition to the effect explained above, with a redshift that depends on the chain length, we find that, depending on both the chain length and the GNR width, the coupling with the chain modes can give rise to different sub-peaks (see e.g. the case of $C_8H_{17}$, Figure 3d). The relaxation of the system symmetry allows for both a mixing of longitudinal (L), transverse (T) and normal (Z) modes, and the activation of modes otherwise forbidden. All of this can explain the broadening of the low frequency peak observed experimentally, which is indeed different for GNRs of different width.

**High-energy spectral region: D and G peaks, overtones and combination modes.** Figure 2b shows the first order Raman spectrum of the ribbons measured at 2.41 eV, with D and G peaks typical of C $sp^2$ materials.[1–6] However, the G peak, which corresponds to the high frequency $E_{2g}$ phonon at $\Gamma$, is up-shifted (~1605 cm$^{-1}$) and broader if compared to pristine graphene (FWHM ~ 25 cm$^{-1}$). Similar results were observed in small graphite domains[35] and PAH[6] due to the relaxation of the momentum conservation induced by finite size. Moreover, we do not observe any splitting of the G peak, as usually found instead for CNTs,[4,5] where this mode, which is doubly degenerate in graphene[36], splits into a longitudinal-optical (LO) and a transverse-optical (TO) component. Our first principles simulations show that the G peak, located at 1618 cm$^{-1}$ for the 8CNR (Figure 3c), is mainly due to the TO mode, with smaller contributions due to TO overtones (within 10 cm$^{-1}$); the LO mode is instead inactive in backscattering, as expected in purely ZZ-GNRs[24]. The presence of the side chains does not alter significantly the nature and position of the main features in the high-energy region of the spectrum, mostly dictated by the edge morphology, as seen by comparing the curves in Figure 3c. Note however that a different position of the functional chain along the GNR edge, as in the case of 6CNR, can induce a distortion of the GNR backbone, thus influencing the peak position (see simulated spectra of 6CNR in Supplementary Information).

Figure 2b also shows a prominent D peak that is characterized by a dominant component at about 1310-1330cm$^{-1}$, with an intensity comparable to that of the G peak, and by one or more shoulders at lower wavenumbers. The Raman shifts of these features are in agreement with our first principles simulations, which show a structured peak in this region, with two main components (at

1336 and 1357 cm$^{-1}$ for the 8CNR, Figure 3c) corresponding to the breathing modes of six-atom rings. This behavior makes the spectrum to resemble that of defective graphene.[37–41] However, there are notable differences that allow us to clearly distinguish our GNRs from defective graphene: the most striking one is found by analyzing the energy dependence of the D peak. In graphene, the D peak – coming from TO phonons around the Brillouin Zone (BZ) edge K— is activated by an inter-valley double resonance process[42] in presence of defects, and it is strongly dispersive with excitation energy due to a Kohn Anomaly at K.[43] The typical D peak dispersion of graphene is ~50 cm$^{-1}$/eV.[44,45] In the case of GNRs, we measure different D-peak dispersions for the different GNRs (see Supporting Information). In particular, the 8CNR, due to its low band-gap (~1.2 eV[30]), allows performing Raman spectroscopy in a wide range of energies (from 1.3 to 2.54 eV) without any photoluminescence background hiding the Raman peaks (Figure 2c). If we fit the dispersion with two slopes, we get about 7 and 35 cm$^{-1}$/eV for the low (< 1.8 eV) and high (> 1.8 eV) excitation energy region, respectively. A considerable dispersion is found also for the RLBM (Figure 2c, bottom panel), while the G peak shows a very small dispersion, at the limit of the spectrometer resolution (Figure 2c, top panel). In addition to this, we observe the G+D combination mode and the 2G mode, not observed in defective graphene. A systematic comparison between defective graphene and our GNRs is reported in the Supporting Information.

To fully understand these experimental observations, one needs to consider that several factors come into play when we move from graphene to narrow GNRs. First, we have neither Kohn anomalies (see calculated phonon dispersion in Supplementary Information) nor linear electronic dispersion in such GNRs, which have semiconducting character, with an optical response dominated by excitonic effects,[46,47] especially for excitation energies close to the optical gap. Second, the K point folds onto Γ in these cove-shaped GNRs:[34] as such, we do not need a double-resonance process to activate this mode, but just a first-order process like in armchair GNRs, where the D peak is also observed.[26] For the same reason, we are also able to observe the G+D mode. In addition, one would expect a non-dispersive behavior for the D peak, similar to the G peak in graphene. The observed dispersion is thus related to some disorder induced scattering, e.g. due to edge functionalization, defects formed during the GNR production, or length distribution.

In conclusion, we combined an experimental and theoretical analysis of the Raman spectra of ultra-narrow, and structurally well-defined graphene nanoribbons with cove-type edges. The low-frequency spectral region contains the main fingerprints of these materials. By analyzing the differences with respect to other systems, such as an ideal ZZGNR and the same GNR without lateral chains (H passivated), ab-initio simulations show that the number of Raman peaks and their position are crucially affected by edge modifications. The full description of cove-type edge and alkyl chains is fundamental to get an agreement with experiments, since both contribute to the shift and splitting of the peaks as well as to a re-distribution of the Raman intensity. The RLBM is especially sensitive, not simply to the width, but also to the edge modulation and functionalization, making it very different from the ideal cases studied to date, where the RLBM does not show significant dependence on edge type. The high-energy spectral region appears similar to that of defective graphene, with D and G peaks of comparable intensity. However, the presence of the D+G combination mode and the different D-peak dispersion allow us to clearly distinguish these GNRs from defected graphene and other graphitic materials.

## Methods

**Experimental details**. Details on the preparation of 4CNR, 8CNR and *m*-ANR are described in Refs. [28–30]. The synthesis of 6CNR will be reported elsewhere. Raman measurements are performed with a combination of different spectrometers (Witec confocal spectrometer, Renishaw InVia and TK64000 by Horiba). A 100x objective is used and the power on the sample is below 0.1 mW to avoid damage. The Raman spectra have been measured with at least three accumulations and in at least 3 different points on the same sample. Measurements were repeated with different spectrometers in different laboratories in different countries to avoid any experimental artifact. The samples are measured as powder, in resonance condition [the optical gap is ~1.9 eV, ~1.2 eV, ~1.12eV for 4-CNR, 8-CNR, and m-ANR, respectively [28-30], while for 6CNR is ~1.8 eV].

**Computational approach.** Simulations are performed using a first-principle plane-wave pseudopotential implementation of Density-Functional Theory (DFT) and Density-Functional Perturbation Theory (DFPT),[48] as available in the Quantum ESPRESSO package.[49] The local density approximation for the exchange-correlation functional is used. Raman intensities are calculated using the second-order response method in Ref. [50], within the Placzek approximation (non-resonant condition). This approach is known to give accurate Raman shifts, while the relative intensity of the peaks cannot be directly compared with resonance Raman experiments. Norm-conserving pseudopotentials are employed, with a plane-wave cutoff energy of 70 Ry. A vacuum region of 12 Å in the non-periodic directions is introduced to prevent interaction between periodic images. The atomic positions are fully relaxed until forces are less than $5 \times 10^{-4}$ a.u. Phonon frequencies and Raman tensor are calculated using a 16×1×1 **k**-point grid.

## Supporting Information

The Supporting Information is available free of charge on the ACS Publications website.

Detailed descriptions of: (I) the effect of the laser power on the first order Raman spectrum of the 4CNR+$C_{12}H_{25}$ ribbon; (II) Multi-wavelength Raman spectroscopy of GNRs as compared to defective graphene; (III) the Raman spectra (experimental and simulated) of all ribbons with different side chains and chain locations; (IV) the zone-folding approximation for cove-shaped GNRs; (V) the D- and G-peak dispersions from DFPT.

## Acknowledgments

We acknowledge funding from: the Alexander von Humboldt Foundation in the framework of the Sofja Kovalevskaja Award, endowed by the Federal Ministry for Education and Research of Germany; the ESF project GOSPEL (Ref. No. 09-EuroGRAPHENE-FP-001); the European Research Council (grant NOC-2D, NANOGRAPH, and Hetero2D); the Italian Ministry of Research through the national projects PRIN-GRAF (Grant No. 20105ZZTSE) and FIRB-FLASHit (Grant No. RBFR12SWOJ); the DFG Priority Program SPP 1459; the Graphene Flagship (Ref. No. CNECT-ICT-604391); the EU project MoQuaS; EPSRC Grants (EP/K01711X/1, EP/K017144/1); the EU grant GENIUS; a Royal Society Wolfson Research Merit Award. Computer time was granted by PRACE at the CINECA Supercomputing Center (Grant No. PRA06 1348), and by the Center for Functional Nanomaterials at Brookhaven National Laboratory, supported by the U.S. Department of Energy, Office of Basic Energy Sciences, under contract number DE-SC0012704.

## Author contributions

I.A.V., M.B., D.Y., and S.M. performed the Raman measurements, with A.C.F. and C.C. providing supervision. M.D.C., A.R., E.M. and D.P. provided the theoretical framework. A.N., Y.H. and M.G.S.

prepared the graphene nanoribbons, under the supervision of X.F and K.M. C.C, D.P. and A.C.F wrote the manuscript, with inputs from all authors.

## Additional information

Competing financial interests: The authors declare no competing financial interests.

## References

(1) Ferrari, A. C.; Basko, D. M. *Nat. Nanotech.* **2013**, *8* (4), 235–246.

(2) Casiraghi, C. In *Spectroscopic Properties of Inorganic and Organometallic Compounds: Techniques, Materials and Applications, Volume 43*; The Royal Society of Chemistry, 2012; pp 29–56.

(3) Jorio, A.; Saito, R.; Dresselhaus, G.; Dresselhaus, M. S. *Raman Spectroscopy in Graphene Related Systems*; Wiley-VCH Verlag GmbH, 2011.

(4) Saito, R.; Dresselhaus, G.; Dresselhaus, M. S. *Physical Properties of Carbon Nanotubes*; Imperial College Press, London, 1998.

(5) Reich, S.; Thomsen, C.; Maultzsch, J. *Carbon Nanotubes: Basic Concepts and Physical Properties*; Wiley-VCH Verlag GmbH, 2004.

(6) Castiglioni, C.; Tommasini, M.; Zerbi, G. *Phil. Trans. R. Soc. Lond. A* **2004**, *362* (1824), 2425–2459.

(7) Ferrari, A. C.; Robertson, J. *Phil. Trans. R. Soc. Lond. A* **2004**, *362* (1824), 2477–2512.

(8) Tuinstra, F. *J. Chem. Phys.* **1970**, *53* (3), 1126.

(9) Ferrari, A. C. *Solid State Commun.* **2007**, *143* (1-2), 47–57.

(10) Tan, P. H.; Han, W. P.; Zhao, W. J.; Wu, Z. H.; Chang, K.; Wang, H.; Wang, Y. F.; Bonini, N.; Marzari, N.; Pugno, N.; Savini, G.; Lombardo, A.; Ferrari, A. C. *Nat. Mater.* **2012**, *11* (4), 294–300.

(11) Lui, C. H.; Malard, L. M.; Kim, S.; Lantz, G.; Laverge, F. E.; Saito, R.; Heinz, T. F. *Nano Lett.* **2012**, *12* (11), 5539–5544.

(12) Wu, J.-B.; Zhang, X.; Ijäs, M.; Han, W.-P.; Qiao, X.-F.; Li, X.-L.; Jiang, D.-S.; Ferrari, A. C.; Tan, P.-H. *Nat. Commun.* **2014**, *5*, 5309.

(13) Wu, J.-B.; Hu, Z.-X.; Zhang, X.; Han, W.-P.; Lu, Y.; Shi, W.; Qiao, X.-F.; Ijäs, M.; Milana, S.; Ji,

W.; Ferrari, A. C.; Tan, P.-H. *ACS Nano* **2015**, *9* (7), 7440–7449.

(14) Rao, A. M.; Richter, E.; Bandow, S.; Chase, B.; Eklund, P. C.; Williams, K. A.; Fang, S.; Subbaswamy, K. R.; Menon, M.; Thess, A.; Smalley, R. E.; Dresselhaus, G.; Dresselhaus, M. S. *Science* **1997**, *275* (5297), 187–191.

(15) Maghsoumi, A.; Brambilla, L.; Castiglioni, C.; Müllen, K.; Tommasini, M. *J. Raman Spectrosc.* **2015**, *46* (9), 757–764.

(16) Di Donato, E.; Tommasini, M.; Fustella, G.; Brambilla, L.; Castiglioni, C.; Zerbi, G.; Simpson, C. D.; Müllen, K.; Negri, F. *Chem. Phys.* **2004**, *301* (1), 81–93.

(17) Terrones, M.; Botello-Méndez, A. R.; Campos-Delgado, J.; López-Urías, F.; Vega-Cantú, Y. I.; Rodríguez-Macías, F. J.; Elías, A. L.; Muñoz-Sandoval, E.; Cano-Márquez, A. G.; Charlier, J.-C. *Nano Today* **2010**, *5* (4), 351–372.

(18) Yazyev, O. V. *Acc. Chem. Res.* **2013**, *46* (10), 2319–2328.

(19) Ferrari, A. C.; Bonaccorso, F.; Fal'ko, V.; Novoselov, K. S.; Roche, S.; Bøggild, P.; Borini, S.; Koppens, F. H. L.; Palermo, V.; Pugno, N.; Garrido, J. A.; Sordan, R.; Bianco, A.; Ballerini, L.; Prato, M.; Lidorikis, E.; Kivioja, J.; Marinelli, C.; Ryhänen, T.; Morpurgo, A.; et al. *Nanoscale* **2014**, *7* (11), 4598–4810.

(20) Zhou, J.; Dong, J. *Appl. Phys. Lett.* **2007**, *91* (17), 173108.

(21) Yamada, M.; Yamakita, Y.; Ohno, K. *Phys. Rev. B* **2008**, *77* (5), 54302.

(22) Vandescuren, M.; Hermet, P.; Meunier, V.; Henrard, L.; Lambin, P. *Phys. Rev. B* **2008**, *78* (19), 195401.

(23) Gillen, R.; Mohr, M.; Thomsen, C.; Maultzsch, J. *Phys. Rev. B* **2009**, *80* (15), 155418.

(24) Gillen, R.; Mohr, M.; Maultzsch, J. *Phys. Status Solidi B* **2010**, *247* (11-12), 2941–2944.

(25) Gillen, R.; Mohr, M.; Maultzsch, J. *Phys. Rev. B* **2010**, *81* (20), 205426.

(26) Cai, J.; Ruffieux, P.; Jaafar, R.; Bieri, M.; Braun, T.; Blankenburg, S.; Muoth Matthiasand Seitsonen, A. P.; Saleh, M.; Feng, X.; Muellen, K.; Fasel, R. *Nature (London)* **2010**, *466*, 470–473.

(27) Dössel, L.; Gherghel, L.; Feng, X.; Muellen, K. *Angew. Chem. Int. Ed.* **2011**, *50* (11), 2540–2543.

(28) Schwab, M. G.; Narita, A.; Hernandez, Y.; Balandina, T.; Mali, K. S.; De Feyter, S.; Feng, X.; Müllen, K. *J. Am. Chem. Soc.* **2012**, *134* (44), 18169–18172.

(29) Narita, A.; Feng, X.; Hernandez, Y.; Jensen, S. A.; Bonn, M.; Yang, H.; Verzhbitskiy, I. A.;

Casiraghi, C.; Hansen, M. R.; Koch, A. H. R.; Fytas, G.; Ivasenko, O.; Li, B.; Mali, K. S.; Balandina, T.; Mahesh, S.; De Feyter, S.; Müllen, K. *Nat. Chem.* **2014**, *6* (2), 126–132.

(30) Narita, A.; Verzhbitskiy, I. A.; Frederickx, W.; Mali, K. S.; Jensen, S. A.; Hansen, M. R.; Bonn, M.; De Feyter, S.; Casiraghi, C.; Feng, X.; Müllen, K. *ACS Nano* **2014**, *8* (11), 11622–11630.

(31) Schwab, M. G.; Narita, A.; Osella, S.; Hu, Y.; Maghsoumi, A.; Mavrinsky, A.; Pisula, W.; Castiglioni, C.; Tommasini, M.; Beljonne, D.; Feng, X.; Müllen, K. *Chem. Asian J.* **2015**, *10* (10), 2134–2138.

(32) Abbas, A. N.; Liu, G.; Narita, A.; Orosco, M.; Feng, X.; Müllen, K.; Zhou, C. *J. Am. Chem. Soc.* **2014**, *136* (21), 7555–7558.

(33) Konnerth, R.; Cervetti, C.; Narita, A.; Feng, X.; Müllen, K.; Hoyer, A.; Burghard, M.; Kern, K.; Dressel, M.; Bogani, L. *Nanoscale* **2015**, *7* (30), 12807–12811.

(34) Osella, S.; Narita, A.; Schwab, M. G.; Hernandez, Y.; Feng, X.; Müllen, K.; Beljonne, D. *ACS Nano* **2012**, *6* (6), 5539–5548.

(35) Ferrari, A. C.; Robertson, J. *Phys. Rev. B* **2000**, *61* (20), 14095–14107.

(36) Basko, D. M. *New. J. Phys.* **2009**, *11* (9), 95011.

(37) Lucchese, M. M.; Stavale, F.; Ferreira, E. H.; Vilani, C.; Moutinho, M. V. O.; Capaz, R. B.; Achete, C. A.; Jorio, A. *Carbon N. Y.* **2010**, *48* (5), 1592–1597.

(38) Martins Ferreira, E. H.; Moutinho, M. V. O.; Stavale, F.; Lucchese, M. M.; Capaz, R. B.; Achete, C. A.; Jorio, A. *Phys. Rev. B* **2010**, *82* (12), 125429.

(39) Cançado, L. G.; Jorio, A.; Ferreira, E. H. M.; Stavale, F.; Achete, C. A.; Capaz, R. B.; Moutinho, M. V. O.; Lombardo, A.; Kulmala, T. S.; Ferrari, A. C. *Nano Lett.* **2011**, *11* (8), 3190–3196.

(40) Eckmann, A.; Felten, A.; Mishchenko, A.; Britnell, L.; Krupke, R.; Novoselov, K. S.; Casiraghi, C. *Nano Lett.* **2012**, *12* (8), 3925–3930.

(41) Eckmann, A.; Felten, A.; Verzhbitskiy, I.; Davey, R.; Casiraghi, C. *Phys. Rev. B* **2013**, *88* (3), 35426.

(42) Thomsen, C.; Reich, S. *Phys. Rev. Lett.* **2000**, *85* (24), 5214–5217.

(43) Piscanec, S.; Lazzeri, M.; Mauri, F.; Ferrari, A. C.; Robertson, J. *Phys. Rev. Lett.* **2004**, *93* (18), 185503.

(44) Ferrari, A. C.; Meyer, J. C.; Scardaci, V.; Casiraghi, C.; Lazzeri, M.; Mauri, F.; Piscanec, S.; Jiang, D.; Novoselov, K. S.; Roth, S.; Geim, A. K. *Phys. Rev. Lett.* **2006**, *97* (18), 187401.

(45) Casiraghi, C.; Hartschuh, A.; Qian, H.; Piscanec, S.; Georgi, C.; Fasoli, A.; Novoselov, K. S.; Basko, D. M.; Ferrari, A. C. *Nano Lett.* **2009**, *9* (4), 1433–1441.

(46) Prezzi, D.; Varsano, D.; Ruini, A.; Marini, A.; Molinari, E. *Phys. Rev. B* **2008**, *77* (4), 1–4.

(47) Denk, R.; Hohage, M.; Zeppenfeld, P.; Cai, J.; Pignedoli, C. A.; Söde, H.; Fasel, R.; Feng, X.; Müllen, K.; Wang, S.; Prezzi, D.; Ferretti, A.; Ruini, A.; Molinari, E.; Ruffieux, P. *Nat. Commun.* **2014**, *5*, 4253.

(48) Baroni, S.; de Gironcoli, S.; Dal Corso, A.; Giannozzi, P. *Rev. Mod. Phys.* **2001**, *73* (2), 515–562.

(49) Giannozzi, P.; Baroni, S.; Bonini, N.; Calandra, M.; Car, R.; Cavazzoni, C.; Ceresoli, D.; Chiarotti, G. L.; Cococcioni, M.; Dabo, I.; Dal Corso, A.; de Gironcoli, S.; Fabris, S.; Fratesi, G.; Gebauer, R.; Gerstmann, U.; Gougoussis, C.; Kokalj, A.; Lazzeri, M.; Martin-Samos, L.; Marzari, N.; Mauri, F.; Mazzarello, R.; Paolini, S.; Pasquarello, A.; Paulatto, L.; Sbraccia, C.; Scandolo, S.; Sclauzero, G.; Seitsonen, A. P.; Smogunov, A.; Umari, P.; Wentzcovitch, R. M. *J. Phys. Condens. Matter* **2009**, *21* (39), 395502.

(50) Lazzeri, M.; Mauri, F. *Phys. Rev. Lett.* **2003**, *90* (3), 036401.

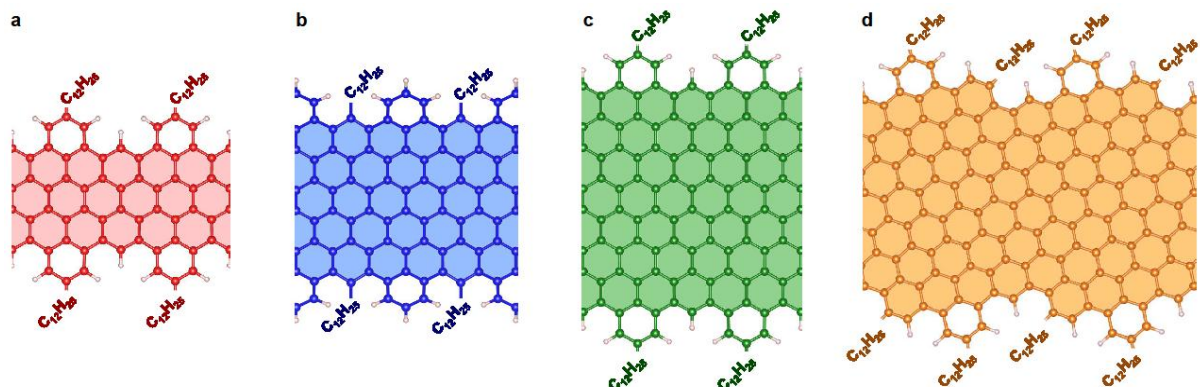

**Figure 1 | Structural model of the GNRs. a-d,** Ball-and-stick representation of the atomic structure of the cove-shaped GNRs investigate here, that is 4CNR (**a**), 6CNR (**b**), 8CNR (**c**), and *m*-ANR (**d**). The schematics show the characteristic variable width of these GNRs and the location of the alkyl side chains in each case. The shaded areas indicate the corresponding zigzag GNRs.

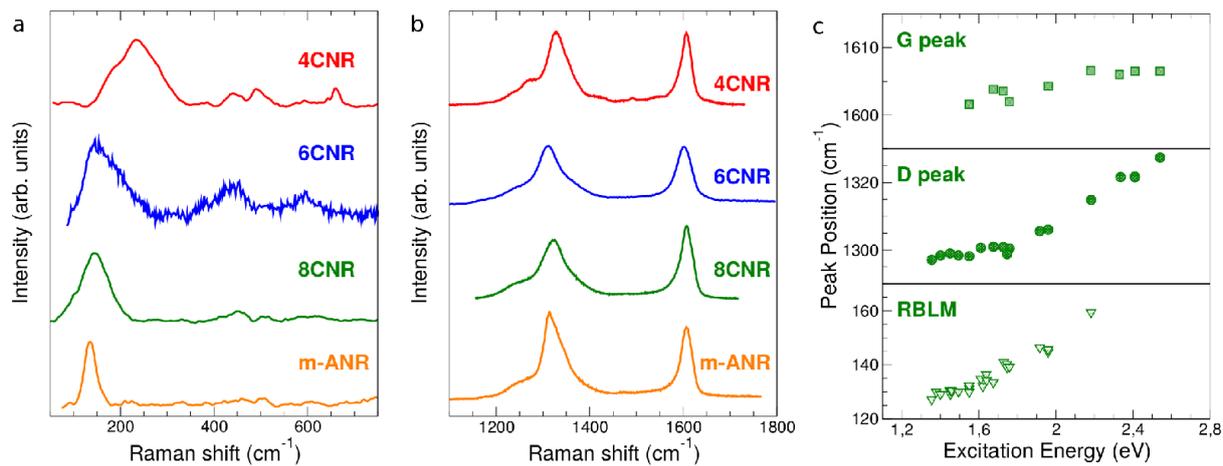

**Figure 2 | Raman spectra of cove-shaped GNRs. a,** Acoustic and **b,** optical region of the Raman spectrum for the cove-shaped GNRs in Figure 1. The 4CNR and 6CNR were excited at ~2.4 eV, while 8CNR and *m*-ANR at ~1.9 eV. **c,** Peak dispersion of 8CNR as a function of excitation energy for the G (top) and D peaks (middle), as well as for the RLBM (bottom).

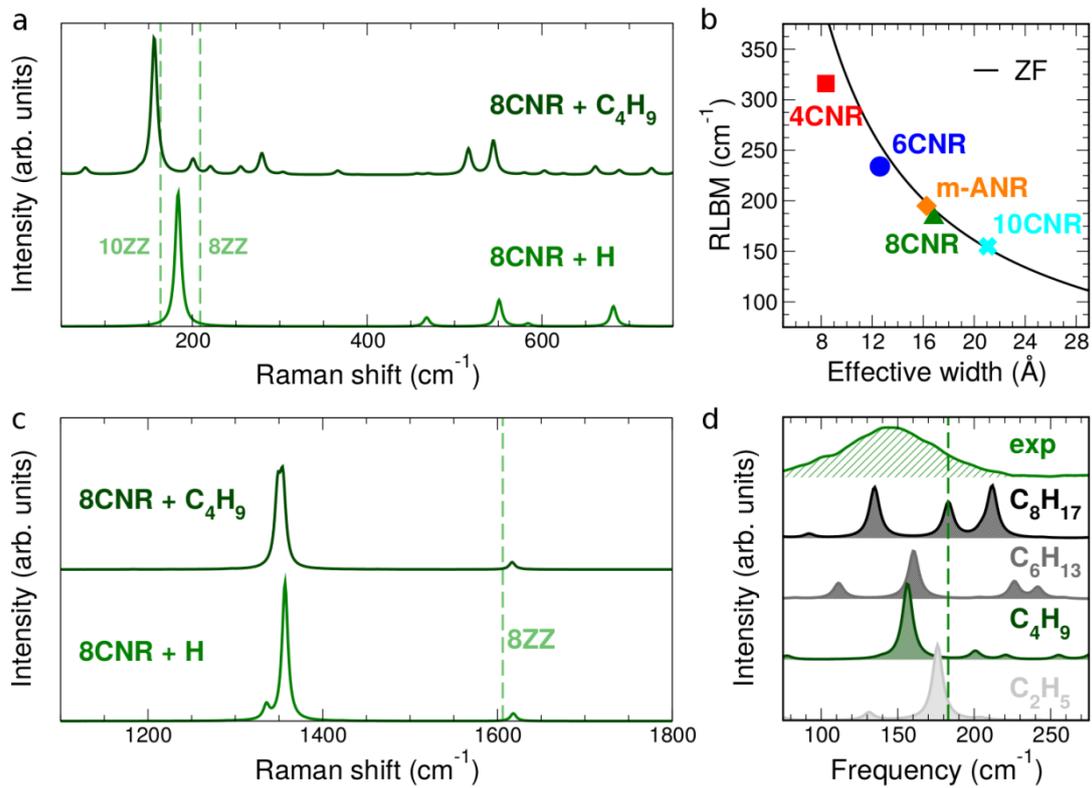

**Figure 3 | Simulated vibrational properties of cove-shaped GNRs. a,** Acoustic and **c,** optical region of the Raman spectrum of 8CNR, as resulting from *ab-initio* DFPT simulations. The spectrum is shown for both hydrogen-terminated (+H, green) and functionalized 8CNR (+$C_4H_9$, dark green). The dashed lines indicate the position of the RLBM for 8- and 10-ZGNRs (labeled 8ZZ and 10 ZZ, respectively, light green) and the position of the G peak for 8-ZGNR. **b,** The frequency of the RLBM calculated from first principles for several H-passivated cove-shaped GNRs is compared to the result of the ZF approximation (black curve) as a function of the GNR effective width. **d,** The low-energy spectral region of the functionalized 8CNR is shown as a function of the chain length, and compared with experimental data. The dashed line indicates the frequency of the RLBM for the H-terminated system. For convenience, a small Lorentzian broadening of ~10 cm$^{-1}$ is introduced in all spectra.